\documentclass[
    ,final            
  ]
  {aipproc}

\layoutstyle{6x9}

\begin{document}

\title{Supernova classes and subclasses}

\classification{97.60.Bw}
\keywords      {Stars: supernovae: general; Stars: fundamental parameters }

\author{Massimo Turatto}{
  address={Osservatorio Astronomico di Padova, vicolo dell'Osservatorio 5, 
  35122 Padova, Italia}
  ,altaddress={KITP, Kohn Hall, UCSB
Santa Barbara, CA 93106, USA} 
}

\author{Stefano Benetti}{
  address={Osservatorio Astronomico di Padova, vicolo dell'Osservatorio 5, 
  35122 Padova, Italia}
  ,altaddress={KITP, Kohn Hall, UCSB
Santa Barbara, CA 93106, USA} 
}
\author{Andrea Pastorello}{
  address={Astrophysics Research Centre, School of Mathematics and Physics, Queen's University Belfast, Belfast
  BT71NN, UK}
}

\begin{abstract}

The discovery of many objects with unprecedented, amazing observational
characteristics caused the last decade to be the most prolific period for the
supernova research. Many of these new supernovae are transitional objects between
existing classes, others well enter within the defined classes, but still show unique
properties. This makes the traditional classification scheme inadequate to take into
account the overall SN variety and, consequently, requires the introduction of new
subclasses.

\end{abstract}

\maketitle


\section{Introduction}
After the appearance of SN~1987A, whose anniversary we are
celebrating, the
number of discovered of Supernovae (SNe) has literally
exploded \citep[][{\it see http://web.pd.astro.it/supern/snean.txt}]{2005ASPC..342...71C}.
The improved monitoring
capabilities, quality of the observations and extension of the
spectral range has produced an unexpected diversification of
the observed properties and, as a consequence, the proliferation of SN types and
subtypes among which only the most experienced specialists are
able to extricate themselves.

The final aim of the SN classification is the
arrangement of these vast arrays of objects into categories according to
basic physical parameters such as initial mass of the progenitor,
explosion mechanism, stellar population, metallicity, etc.
To this aim all kinds of information should be used including, in
addition to the conventional data on the SN
(spectroscopy, photometry, polarimetry), also information on the parent
population (locations inside the parent galaxy, stellar population at the
site of explosion, pre-explosion observations and circumstellar environment) 
as well as particle flux (neutrinos, cosmic rays).

Usually a {\it quick and dirty} SN classification (typing) based on the
presence/absence of spectral features is obtained as soon as possible
after the discovery in the framework of {\it ad hoc} projects on small/medium
class telescopes which make use of their flexible scheduling to
accommodate Target of Opportunity observations. This quick
classification, required to point out the most interesting
targets for subsequent intensive observational campaigns, usually
hold extensive studies. On the other hand there are well studied
objects whose physical nature is debated also after long--standing 
campaigns, e.g. SN~2002ic (cfr. next Sections).

High redshifts SNe are nowadays discovered on-demand by patrolling large
volumes of Universe with wide-field imagers mounted on large telescopes. Their
faintness prevents extensive studies on individual objects
and the type assignment for all candidates. In the wake of wide-field
spectroscopic multiplexer on large telescopes \citep[e.g.][]{2004SPIE.5492..121R},
statistical methods have been developed which make
use of the SED distribution derived from the
photometry to type the SN candidates with a reasonable probability
\citep{2006astro.ph.10129P}.

Aim of this review is to update the general classification scheme
of SNe on the basis of the latest results also discussing the
most recent objects challenging the consolidated scenario.
Detailed classification schemes for SNe can be found in
\citep{1997ARA&A..35..309F,HarkWheel,WheelBen,2003LNP...598...21T}.

\section{The Consolidated Scenario}
 
It is widely accepted that SNe result from two major explosion
mechanisms, the gravitational collapse of the stellar nucleus (core collapse) and the 
thermonuclear runaway. The core collapse
takes place in massive stars ($M\geq 8 M_\odot$) at the end of a series
central nuclear burnings which end up with the formation of an iron core,
and results in the formation of a compact remnant, a neutron star (NS) or a
black hole (BH). The different
configurations of the progenitors at the moment of the explosion, the
different energetics associated to the event and the possible interaction
of the ejecta with circumstellar material (CSM) 
produce a large variety of displays.

The thermonuclear disruption of CO White Dwarfs (WDs), which have reached
the Chandrasekhar mass limit accreting matter from a companion in binary
systems, produces those we normally call type Ia Supernovae (SNIa). Whether
the donor is another degenerate star or a main sequence/red giant star
is still debated, as well as whether the burning front propagates as a
subsonic deflagration or turns into a detonation, in the so-called 
delayed--detonation models \citep{2000ARA&A..38..191H}. Recent 3-D
deflagration models seem to fail in representing true SNIa explosions,
both because the low energy released in this kind of explosion and the full
mixing between fuel and ashes not seen in SNIa spectra 
\citep{2003Sci...299...77G}.

\section{Type Ia supernovae}

The CO WD scenario well explains the main observational properties of
these objects such as the lack of H lines, the occurrence in all type of
galaxies and their overall similarity. Nevertheless in the nineties a new
paradigm for SNe Ia was developed. Intrinsic differences were identified
in the luminosities at maximum which correlate with the early light curve
shape, brighter objects having broader light curves than dimmer ones
\citep{1993ApJ...413L.105P,1996ApJ...473...88R,1997ApJ...483..565P,
1999AJ....118.1766P,2006ApJ...647..501P}. 
This sequence was matched by spectroscopic
variations attributed to changes in the effective temperatures, which, in the
context of Chandrasekhar-mass explosions, were interpreted as variations
in the mass of $^{56}$Ni produced in the explosions 
\citep{1995ApJ...455L.147N}.
These correlations have been employed in restoring SNIa as useful distance
indicators up to cosmological distances but do not fully account for the entire
dispersion/diversity observed \citep[e.g.][]{2004MNRAS.348..261B}.

The Chandrasekhar mass scenario has been recently challenged by SN~2003fg, 
a bright (2.2 times more luminous than normal) SNIa 
with usual spectral features and relatively low
expansion velocity. SN~2003fg did not obey the empirical relations between the light curve shape
and the luminosity at maximum and therefore objects like this,
though intrinsically very rare in the local Universe, may 
contaminate the searches of high-z SNIa for cosmology.
The progenitor was estimated to be 2.1 M$_\odot$ (i.e. {\bf SuperChandra})
\citep{2006Natur.443..308H,2006astro.ph..9804J} 
which in principle might form by the merge of two massive WDs. 
However also the possible interpretation with a Chandrasekhar mass
model has recently been proposed \citep{2007A&A...465L..17H}.


Based on the analysis of the photometric and spectroscopic properties of a
sample of 26 objects \citet{2005ApJ...623.1011B} have identified three
subclasses of SNIa with distinct physical properties, the main characterizing
parameter being the gradient of expansion velocity of the photosphere. {\bf
Faint} SNIa (similar to SN~1991bg) are fast decliners both in  luminosity
and expansion velocity, have typically low expansion velocities and occur in
early-type galaxies. High-- ({\bf HVG}) and low--velocity ({\bf LVG}) 
gradient SNIa include normal objects, although the LVG group also includes all the
brightest, slow declining SNe (like SN 1991T). Even if the statistical
analysis suggests that LVG and HVG are two distinct groups, they may
possibly represent a continuum of properties.
More peculiar
objects, like SNe 2000cx \citep{2001PASP..113.1178L} and 2002cx 
\citep{2003PASP..115..453L,2007PASP..119..360P} were not included in 
the above analysis.

In a recent work \citet{2007Sci...315..825M} have performed a
systematic spectral analysis of a large sample of well observed type Ia
SNe (including SN~2000cx). Mapping the velocity distribution of
the main products of nuclear burning, they tried to constrain the
theoretical scenarios behind the SNIa phenomena. They found that all
supernovae have low-velocity cores of stable Fe-group elements, while
outside radioactive $^{56}$Ni dominates the supernova ejecta. The outer extent of
the iron-group material depends, thus, on the amount of $^{56}$Ni and
coincides with the inner extent of Si, the principal product of
incomplete burning. Surprisingly, they found that the outer extent of the
bulk of Si is similar in all SNe ($\sim 11000$ km s$^{-1}$)
corresponding to a mass slightly over 1M$_\odot$. This indicates that
all considered SNe burned a similar amount of material, and this
suggests that their progenitors had the same initial mass. Synthetic light
curve parameters and three-dimensional explosion simulations support this
interpretation. A single explosion scenario, possibly a delayed detonation,
may thus explain most SNIa. What then drives the main explosion outputs 
(e.g. luminosity) is the strength of the central deflagration and the extent of 
star burned before the detonation goes off. The remaining diversity among SNIa
could then be explained by 3-D effects present in the outer part of the envelopes
like ejection of blobs of burned material along the line of sight and
interaction with the CSM. These phenomena are also responsible for the high
velocity features (HVFs) seen in early spectra \citep{2005ApJ...623L..37M}.

\section{Type Ia supernovae and CSM}

In the past years many attempts aimed to identify the progenitor systems
of SNIa have been carried out, some of them focused on the  
detection and possible characterization of the CSM. 
The radiation arising
from the interaction between the fast moving SN ejecta and the slow moving
CSM in the form of narrow optical emission lines has been searched for 
\citep{1996MNRAS.283.1355C,2005A&A...443..649M}, as well as
radio \citep{2006ApJ...646..369P} and X-ray emission 
\citep{2006ApJ...648L.119I}. 

Radio observations
provide very stringent limits which in a few cases are
as low as $3\times 10^{-8}$ M$_{\odot}$ yr$^{-1}$ for an assumed wind
velocity of 10 km s$^{-1}$ \citep{2006ApJ...646..369P}. 
Recently, the sub-luminous SN 2005ke has been tentatively detected in X-rays domain
by mean of deep monitoring with the XRT onboard of Swift satellite. The inferred X-
ray luminosity [L = ($2\pm 1)\times 10^{38}$ erg s$^{-1}$ in the 0.3--2 keV band]
has been interpreted as the interaction of the SN ejecta with circumstellar
material deposited by a stellar wind from the progenitor's companion star having
a mass-loss rate of $3\times 10^{-6}$ M$_\odot$ yr$^{-1}$ \citep{2006ApJ...648L.119I}.

A more robust detection of CSM around SNIa has been obtained
using high resolution spectroscopy.
Repeated observations of the interstellar/circumstellar NaID doublet in the normal
type Ia SN~2006X have revealed short--scale time variations \citep{patat2006X}. The
expansion velocities, densities and dimensions of the CS envelope indicate that this
material was ejected from the progenitor system short before the explosion (50 years).
Among others clues, the relatively low expansion velocities ($\sim 50$ km s$^{-1}$)
favor a progenitor system where, at the time of the explosion, the WD is accreting
material from a non-degenerate (probably a red-giant star) companion. Moreover, the
amount of CS material deduced (a few $10^{-4}$ M$_\odot$) is compatible with the radio
non-detection of normal SNIa.

In the context of evanescent CSM around SNIa two remarkable exceptions are
represented by SNe 2002ic and 2005gj, which have shown pronounced H emission
lines
\citep{2003Natur.424..651H,2006ApJ...650..510A} that have been
interpreted as a sign of extremely strong ejecta-CSM interaction. It must be noted, 
however, that their
classification as interacting type Ia SNe has been recently been
questioned by
\citet{2006ApJ...653L.129B}, who suggested that SN~2002ic (and SN~2005gj
\citep{Benetti_2005gj}) actually were type
Ic SNe, i.e. that their progenitors were massive. 
In this scenario the observed interaction
with dense circumstellar material is the predictable consequence of the
intense mass-loss activity of the progenitor. 
Because of the striking similarity of the late--time spectra with those
of SNe 1997cy and 1999E, SN~2002ic 
establishes a link between energetic SNIc and the highly interacting SNIIn,
and add some credit to the proposed association of these SNIIn to GRBs
\citep{Germany00,Turatto00,Rigon03,2007arXiv0705.3057B}. 

If the scenario proposed 
for SNe 2002ic and 2005gj is correct, the detection of
CSM in SN~2006X \citep{patat2006X} remains the only certain signature 
of CSM around the progenitor system of a SNIa.

\section{Type II supernovae}

Type II SNe represent core-collapse explosions occurring in progenitor
stars still retaining the H envelopes. They span large ranges in the
observables like luminosity at maximum, explosion energy,
ejected Ni mass, etc. \citep{1997ARA&A..35..309F,1993A&AS...98..443P}. 
Three major classes are often considered, type II Plateau SNe
({\bf SNIIP} \citep{1979A&A....72..287B}) with flat 
light curves in the first
few months, type II Linear SNe({\bf SNIIL} \citep{1979A&A....72..287B}) with 
rapid, steady decline in the
same period, and the narrow-lined SNII ({\bf SNIIn} \citep{1990MNRAS.244..269S}), 
dominated by emission
lines with narrow components, sign of energetic interaction between
the SN ejecta and the CSM. 
Numerous intermediate objects among SN classes exist as well as a large variety 
within each individual class. 
Several attempts have been made to sort out more physical
classification schemes although none of them has proved to be fully
satisfactory \citep[e.g.][]{1994A&A...282..731P}.

\begin{figure}
  \includegraphics[height=.5\textheight]{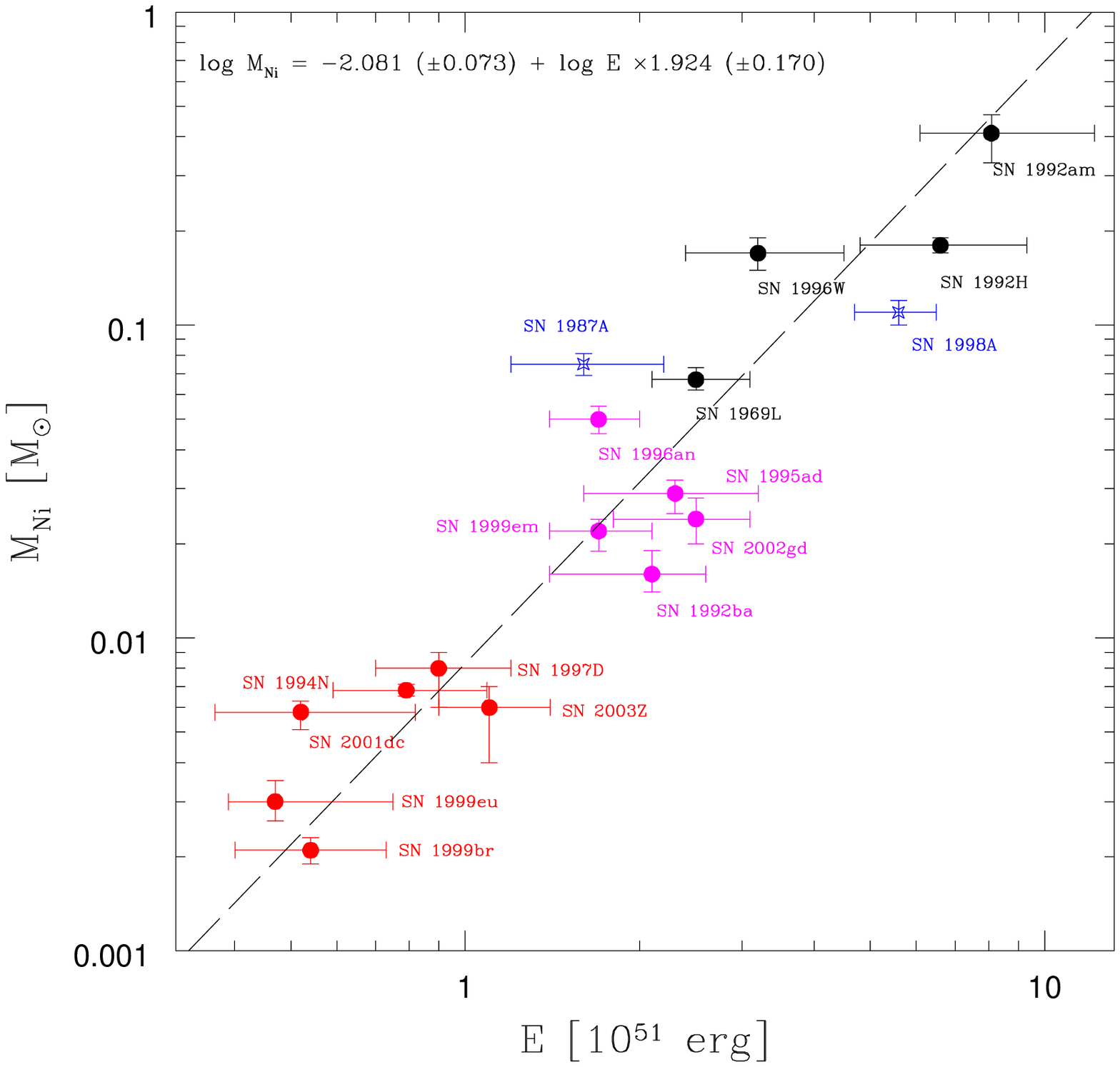}
  \caption{Ejected $^{56}$Ni mass vs. explosion energy (E)
  for a sample of SNIIP \citep[from][]{Pasto_PhD_Thesis}.}
  \label{fig_pasto}
\end{figure}

SNIIP have been the subject of extensive analysis by \citet{2003ApJ...582..905H} 
who pointed out a continuum in the properties of these objects and 
revealed several relations linking the observables and the derived physical
parameters, with more
massive progenitors producing more energetic explosions. 
A similar, independent analysis \citep{Pasto_PhD_Thesis} has been performed
on a sample more extended toward the low-- and high--luminosity objects,
confirming the continuity in the physical properties over a very broad
range. In particular, the group of {\bf faint,
slowly expanding} SNe with characteristics similar to SN~1997D
\citep{1998ApJ...498L.129T,2004MNRAS.347...74P}, which are 
explained by spherical explosions that undergo a large amount of
fallback, appears as the low luminosity tail of the distribution of properties 
of SNIIP. 
SN~1987A and the somehow similar SN~1998A \citep{2005MNRAS.360..950P},
having light curves with
pronounced secondary maxima at about 3 months after the explosion,
are two outliers in the SNIIP family.
In Fig.~\ref{fig_pasto} is shown the $^{56}$Ni mass as a function of the explosion
energy. These quantities correlates very well over two orders of magnitudes in
M$_{Ni}$ and one in E.

The existence of {\bf ultrafaint} (still undetected) core--collapse SNe, 
even fainter than SN~1997D, has been suggested in the framework of GRBs. 
Indeed there are at least two examples of nearby, long-duration GRBs, for which no
optical counterpart brighter than M$_V=-13.5$ has been
identified \citep{2006Natur.444.1050D,2006Natur.444.1047F,2006Natur.444.1053G}. 
\citet{2007ApJ...657L..77T} have shown that faint (and ultrafaint), low-energy SNe with little $^{56}$Ni
ejection are compatible with relativistic jet-induced BH-forming
explosions of massive stars.

\section{Type Ib/c supernovae}

The core--collapse of massive stars which have lost their H (and He)
envelopes before the explosion produces SNIb (SNIc).
These objects have recently deserved large attention because of their
relation to GRBs. Indeed long-duration GRBs at sufficiently close
distance have been coupled to bright, highly energetic type Ic SNe.

SN~2003dh associated to GRB030329 \citep[e.g.][]{2003ApJ...591L..17S}
and SN~2003lw/GRB031203 \citep[e.g.][]{2004ApJ...609L...5M} were similar to the
first protypical SN~1998bw/GRB980425 having broad--line
SNIc features. Detailed analysis showed that these SNe require even more
than $10^{52}$ erg, somewhat less in case of asymmetric explosion, 
justifying the introduction of the term {\bf hypernovae} \citep{1998Natur.395..672I}.
They are believed to be the outcome of very energetic
black hole forming explosions of massive stars (30-50 M$_\odot$) which
synthesize large amounts of $^{56}$Ni (0.3-0.5 M$_\odot$).
In addition to the above mentioned cases, there are a number of {\bf broad--line 
SNIc} for which an accompanying GRB has not been detected, e.g. SNe 1997ef
\citep[e.g.][]{2000ApJ...534..660I}, 2002ap \citep{2002ApJ...572L..61M}, 
2003jd \citep{Valenti_2003jd}.
These SNe have a tendency to have smaller luminosity, mass of the ejecta and
explosion energy than GRB--SNe but it is not
clear whether the non-detection of the GRB is a geometric effect
due to asymmetries or an intrinsic property of the explosion.

The soft, nearby GRB060218, classified as an X-Ray Flash (XRF),
was associated to another SNIc, 2006aj, which was less bright than 
GRB--SNe and whose spectral lines were not as broad \citep{2006Natur.442.1011P}.
The spectral and light curve modeling indicate that the mass of the
ejecta is smaller (2 M$_\odot$), the amount of $^{56}$Ni  still considerable
(0.2 M$_\odot$),  and the explosion energy only marginally
larger than the canonical $10^{51}$ erg. 
According to \citet{2007astro.ph..2472N} the considerable variety in the 
broad--line SNIc (associated to GRBs, XRF or not associated) may be explained 
in a unified scenario with jet--induced explosions.

\begin{figure}
  \includegraphics[height=.5\textheight]{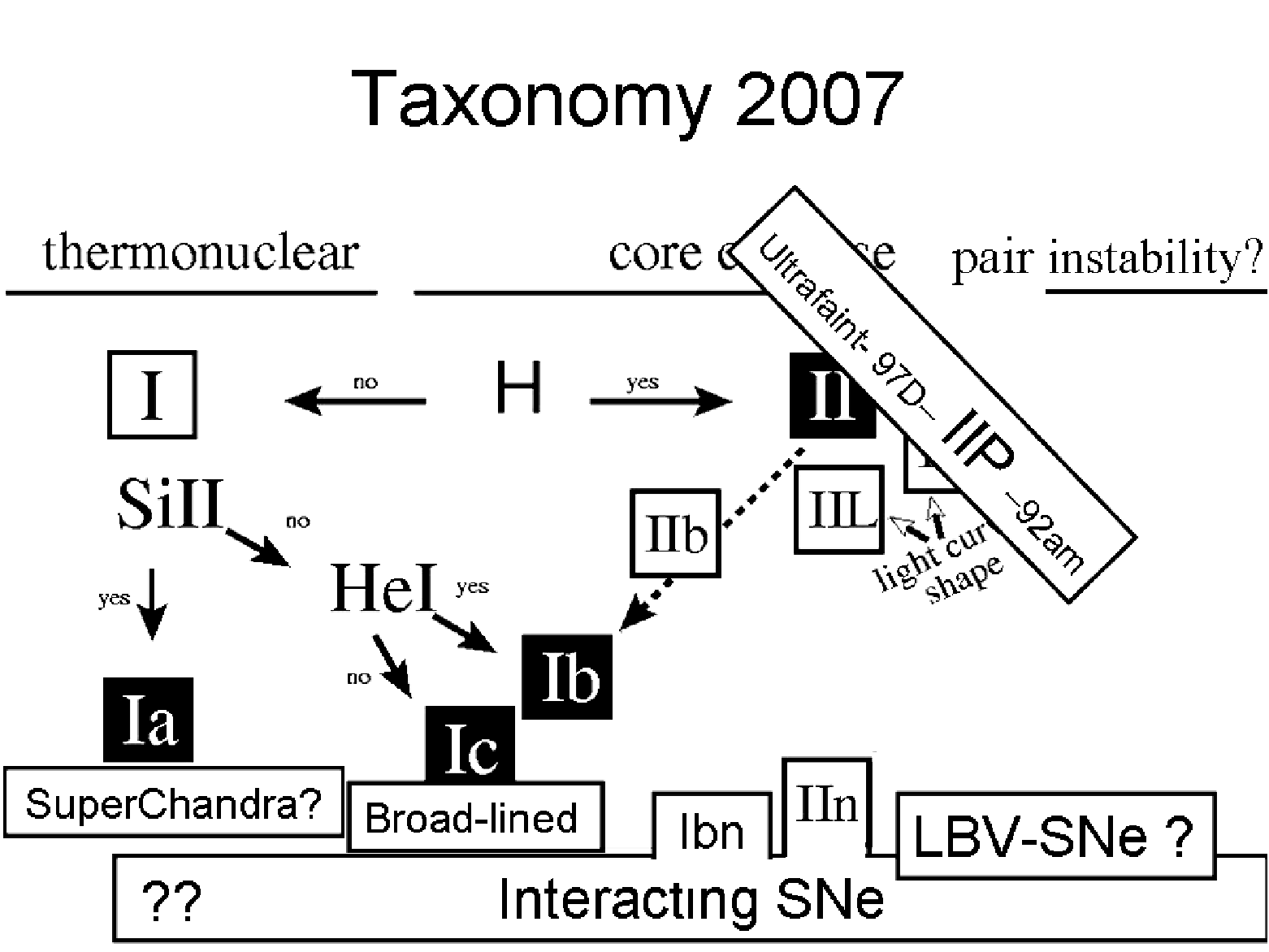}
  \caption{The 2007 version of the classification scheme of SNe.
  It should be considered as an updated version of Figure 1 of 
  \citet{2003LNP...598...21T}.}
  \label{newtaxo}
\end{figure}

\section{Supernovae from extremely massive progenitors}

In the last months, before the beginning of this conference, two remarkable SNe have
occurred which might shed light on the final fate of extremely massive stars.

SN~2006gy was discovered close to the nucleus of the parent galaxy. For this
reason it was early debated if it was actually a SNIIn \citep{2006CBET..647....1H} or an
AGN \citep{2006CBET..648....1P}. Further observations proved that it was
a SN with
unique characteristics: very slow rise to maximum (about 70d) and decline, coupled
to an extremely bright luminosity, M$_{V, max}\sim-22$ 
\citep{2007ApJ...659L..13O,2006astro.ph.12617S}. 
These make the total radiated energy larger than $10^{51}$ erg,
which can be explained only with the explosion of a very massive stars either if
the light curve is powered by CSM interaction or by radioactive decay of 
$^{56}$Ni \citep{2006astro.ph.12617S}. The presence of several solar masses of unshocked
circumstellar gas expanding at 130-260 km s$^{-1}$ is inconsistent with masses of
the progenitor smaller than 40 M$_\odot$. Therefore
\citet{2006astro.ph.12617S} have suggested that 
it was the explosion of a LBV star, like $\eta$ Carinae, with an
initial mass as large as 100-150 M$_\odot$, which failed to shed its entire H
envelope before the explosion and did not become a WR star.

The explosion mechanism invoked would, then, be the pair--instability
collapse, for which 
theoretical light curves have been presented \citep{2005ApJ...633.1031S}.
Indeed pair-production SNe are expected to produce very broad
light curves with bright maxima, to have spectra showing H lines with 
expansion velocity of the order of 5000 km s$^{-1}$, and
to synthesize up to several tens of solar mass of $^{56}$Ni.
Although these models have been computed only for non-rotating stars
of zero metallicity, the pair-production SNe hypothesis is certainly an intriguing
scenario. If confirmed, this unique SN might reveal that some very 
massive stars do not collapse directly onto a BH but explode with 
a bright display with obvious consequences on the possible detection
of Population III SNe.

Shortly after SN~2006gy another object, SN~2006jc, appeared with optical display
very different but still astonishing. 
The spectrum was that of an H--poor event with
broad  SNIc features, much narrower He emissions and very blue continuum 
\citep{2007astro.ph..3663P,2007ApJ...657L.105F}. 
For this reason it has been proposed \citep{2007astro.ph..3663P} that
it should be more properly classified as a {\bf type Ibn} SN. 
The narrow He I spectral emissions together with the presence of a 
strong X--ray emission were evidences of interaction between the ejecta 
of an CO Wolf--Rayet star with a dense He--rich CSM. 
Despite these peculiarities, the most interesting property of SN~2006jc 
is that it was spatially coincident with a bright
(M$_R=-14.1$) optical transient detected in 2004 
\citep{2007astro.ph..3663P}. Similar outbursts
characterize LBVs, massive stars showing significant optical variability due to
unstable H--rich atmospheres and episodic mass loss.
Current evolutionary
theories do not predict the core collapse during the LBV phase but 
statistical arguments strongly disfavour the random spatial coincidence 
of these two events.
It appears therefore that we have testified
in 2004 a sort of LBV--like eruption of an WR, which subsequently exploded as
SN~2006jc \citep{2007astro.ph..3663P,2007ApJ...657L.105F}.

\section{Conclusions}

The current classification scheme of SNe is an evolving tool scientists have
developed to sort out  groups of objects physically related. 
Despite the long record of SNe discovered and studied so far 
\citep[][and on--line updates]{1999A&AS..139..531B} the taxonomy of SNe is far
from being steadily defined and understood. Not only we are still discovering new
common properties and differences among the objects of consolidated types, but
also we are still identifying new classes which are opening new unexpected
scenarios on the final fate of stars.
In Fig.~\ref{newtaxo} we report an updated version of the SN taxonomy
published in \citet{2003LNP...598...21T} to the light of the most recent 
findings discussed above.


\begin{theacknowledgments}
MT and SB acknowledge KITP for the hospitality during the writing of this paper.
As such, this research was supported in part by the National Science Foundation
under Grant No. PHY05-51164.
This research has been supported by the MURST-PRIN n.2006022731.
\end{theacknowledgments}



\bibliographystyle{aipproc}   

\bibliography{turatto_Aspen_fin}

\IfFileExists{\jobname.bbl}{}
 {\typeout{}
  \typeout{******************************************}
  \typeout{** Please run "bibtex \jobname" to optain}
  \typeout{** the bibliography and then re-run LaTeX}
  \typeout{** twice to fix the references!}
  \typeout{******************************************}
  \typeout{}
 }

\end{document}